\documentclass{jetpl} \twocolumn \lat

\def\be{\begin{equation}}
\def\ee{\end{equation}}
\def\bea{\begin{eqnarray}}
\def\eea{\end{eqnarray}}
\def\ba{\begin{array}}
\def\ea{\end{array}}
\def\tr{\mathop{\rm tr}}
\def\Teff{T_e}
\def\Th{E_{Th}}

\newcommand{\sign}{\mathop{\rm sign}\nolimits}
\newcommand{\e}{\epsilon}
\newcommand{\w}{\omega}
\newcommand{\G}{\Gamma}
\newcommand{\p}{\partial}
\newcommand{\kp}{\kappa}
\newcommand{\g}{\gamma}

\title{AC Josephson effect in the long voltage-biased SINIS junction}
\rtitle{AC Josephson effect\ldots} \sodtitle{AC Josephson effect in
the long voltage-biased SINIS junction}
\author{K.\,S.\,Tikhonov$^{+,*}$\/\thanks{e-mail: tikhonov@itp.ac.ru},
M.\,V.\,Feigel'man$^{*,+}$}
\rauthor{K.\,S.\,Tikhonov, M.\,V.\,Feigel'man} \sodauthor{Tikhonov,
Feigel'man}
\address{$^+$Moscow Institute of Physics and Technology, Moscow 141700, Russia
\\
~
\\
$^*$L.D.Landau Institute for Theoretical Physics RAS, 117940 Moscow,
Russia}
\dates{15 January 2009}{25 January 2009}

\abstract{Theory of non-stationary coherent effects is developed for
superconductor-normal-superconductor (SNS) structures with
relatively strong normal scattering on S/N interfaces (interface
resistance  is large compared to intrinsic resistance of N metal).
Analytical expressions are found for the time-dependent anomalous
Green functions induced in the N region under the
fixed-voltage-bias. The amplitude of the current oscillations is
determined in
 non-equilibrium conditions.
Non-stationary correction to the distribution function is calculated
in high-temperature limit and found to be slowly decreasing with the
temperature, leading to the dominance of the second-harmonic term in
the Josephson current, $I_s(t) \propto \sin(4eVt)$ at high
temperatures and low voltage.}

\begin{document}
\maketitle
\noindent
{\bf 1. Coherence effects and general equations.} \\
The superconducting hybrid (normal metal - superconductor)
structures are very rich systems that have been studied for the past
few decades both theoretically and experimentally (cf. 
~\cite{review} for a relatively recent brief review). The proximity
effect has been shown to induce superconductive correlations into
normal part of the system, where they decay over the length
$\xi_\e=\sqrt{{\hbar D}/{\e}}$ that can be large at low energies.
One of the consequence of the proximity effect is the ability of the
normal metal to carry out phase-sensitive current if length $L$ of
the junction is smaller then $\xi_{\e_0}$ at characteristic energies
$\e_0$.  Equilibrium Josephson effect in diffusive SNS systems is
well understood by now and can be well explained in terms of
stationary Andreev levels or more directly through Green's functions
approach. When the constant voltage is applied to the junction,
nonstationary Josephson effect arises. If the voltage is high,
$V\gg\Th$, than effects of coherence between superconducting
reservoirs can be neglected and current is stationary with the rich
subgap structure \cite{Bezuglyi}. When the voltage is not very high,
time-dependent contribution to the current can become
important, even at rather high temperatures~\cite{Lempitsky,Argaman,Brinkman,Zhou}.

In this paper we consider nonstationary Josephson effect in a long
symmetric voltage-biased SNS junction with large interface
resistance, so that $r={R_B}/{R_N}\gg 1$ (here $R_B$ is the single
barrier resistance and $R_N$ the resistance of the normal wire).
This problem was first considered in \cite{LarkinAslamazov}, where Josephson
current up to the first order in $r^{-1}$ was calculated. It was
shown, that Josephson current decays exponentially when temperature
is high comparative to the inverse diffusion time. However,  it was shown later
\cite{Argaman}, that due to
nonequilibrium effects time-dependent current does decay  as slow as $T^{-1}$ at high temperatures.
Here we microscopically calculate the current, with all
nonequilibrium effects taken into account. In doing so, we have to include the terms
which are formally of higher orders in $r^{-1}$; however the resulting current
is not necessary smaller than lowest-order term, but can dominate it,
as explained below.

To describe this system we use Keldysh method, developed for
superconductivity by Larkin and Ovchinnikov~\cite{LO1}. In this
method the Green function is $4 \times 4$ matrix in Keldysh and
particle/hole $2\times 2$ spaces. This matrix $\check G$ contains
all the information about spectrum of the system and on the
distribution of electrons over the energy levels. The resulting
Usadel equation is written in terms of the disorder-averaged
semiclassical Green function $\check G=\check G(t_1,t_2,{\bf r})$
(for the detailed review, see \cite{Rammer}). This equation is
presented below, it can be used to describe any nonstationary
phenomena with low energy scales involved (compared to Fermi
energy). It involves time convolution operation: $\left(f\circ
g\right)(t_1, t_2)=\int_{-\infty}^\infty f(t_1, t)g(t, t_2)dt$. It's
convenient to introduce $t=(t_1+t_2)/2,\;\;\tau=t_1-t_2$. Then
Usadel equation in the absence of the vector potential (which is
supposed to be zero throughout the paper) and electron pairing reads
as follows (we work in units $e=k_B=\hbar=1$): \be \label{Usadel}
-D\p_x(\check G\circ\p_x\check G)+\p_\tau[\check \sigma^3, \check
G]+\frac12\p_T\{\check \sigma^3, \check G\}+i\varphi_-\check
G=\check I \ee with $\check\sigma^3=\check 1\hat\tau^3$ and $\check
I=-i\left(\check\Sigma_{in}\circ\check G-\check
G\circ\check\Sigma_{in}\right)$ where $\check\Sigma_{in}$ is
self-energy matrix, that accounts for inelastic and dephasing
processes. Time-dependent electric potential $\varphi(t)$ enters
this equation through $\varphi_-(t_1,
t_2)=\varphi(t_2)-\varphi(t_1)$. It has to be determined
self-consistently via electroneutrality condition
 $\varphi(t)=\frac{\pi}{4}\tr\hat G^K(t,t)$ \cite{Rammer}.

Keldysh Green function $\check G$ obeys normalization condition
$\check G \circ \check G=\check 1\delta(t_1-t_2)$ which allows for
the anzatz $\hat G^K=\hat G^R\circ\hat h - \hat h\circ\hat G^A$
where $\hat h$ stays for the matrix distribution function. The
latter  can be chosen to be diagonal: $\hat h =
h_0\hat\tau^0+h_3\hat\tau^3$.
It is convenient to implement Fourier transform over $\tau$, i.e. to
pass to  Wigner representation. This way, we write $\check
G(\tau,t)=\int \check G(\e,t)e^{-i\e\tau}d\e$. We note that the
convolution operator simplifies in the mixed representation due to
the following identity: for $f(\e, t)=e^{-i\w_1 t}f(\e),\;\;g(\e,
t)=e^{-i\w_2t}g(\e)$, one gets
$$
\left(f \circ g\right)(\e,t) = e^{-i(\w_1+\w_2)t}
f(\e+\frac{\w_2}{2}) g(\e-\frac{\w_1}{2})
$$
Our goal is to find $\hat G^R,\;\hat G^A, \hat G^K$ for the SINIS
structure and to calculate the current. Current density is expressed
through Keldysh component of the  matrix current $\check j=\check
G\circ\nabla\check G$ as follows: \be \label{Current}
I(t)=\frac{\pi\sigma_N}{4}\tr\hat \tau^3 \hat j^K(t,t) \ee
Matrix equation
(\ref{Usadel}) has to be supplemented with the
Kupriyanov-Lukichev~\cite{KLpaper} boundary condition at both SIN
interfaces. At the right interface it takes the form: \be
\label{Kupriyanov} 2R_{SN}\sigma_N \check j= [\check G\circ, \check
G_r] \ee with $R_{SN}$ being the interface resistance of the barrier
per unit area in the normal state. We use indices $l,r$
for the left and right reservoirs.

Now we have to work out retarded, advanced and Keldysh components of
equations (\ref{Usadel},\ref{Kupriyanov}). In the explicit form,
they are presented in the Appendix. Retarded and advanced components
determine the generalized spectral properties of the system and
Keldysh components describe distribution of electrons over this
generalized spectrum. In what follows, we  simplify these equations
using the smallness of the parameter $r^{-1}$, as described in the
next section.

We consider the constant-voltage-biased setup and neglect the
spatial dependencies of all the relevant quantities in the
directions, perpendicular to the wire. We measure the length in
units of  $L$, assuming that bulk superconductors are situated at
points $x=-\frac{1}{2}$ and $x=\frac{1}{2}$ at the voltages
$-\frac{V}{2}$ and $\frac{V}{2}$ correspondingly. We suppose that
bulk superconductors are good reservoirs at temperature $T_S$; under this assumption their
Green functions $\check G_{l,r}$ can be obtained from the standard
BCS form $\check G_{BCS}$ by the gauge transformation $\check
G_{l,r}(t_1, t_2)=\check S_{l,r}(t_1)\check G_{BCS}(t_1,t_2) \check
S_{l,r}^+(t_2)$ with $\hat
G_{BCS}^{R(A)}=\left(g_S\hat\tau^3+f_S\hat\tau^1\right)^{R(A)},\;\hat
G_{BCS}^K=\tanh(\frac{\e}{2T_S})\left(\hat G_{BCS}^R-\hat
G_{BCS}^A\right)$ and $\check S_{l,r}(t)=
\exp(\pm\frac{iVt}{2}\hat\tau^3)\check 1$.
Straightforward calculation gives:
\begin{equation}
\hat G_{l,r}^R(t,\e)=\left(
\begin{array}{cc}
g_S^R(\e\mp V/2)&e^{\mp iVt}f_S^R(\e)\\e^{\pm
iVt}f_S^R(\e)&-g_S^R(\e\pm V/2)
\end{array}
\right)
\end{equation}
and
\begin{equation}
\hat h_{l,r}(\e)=\left(
\begin{array}{cc}
\tanh(\frac{\e\mp V/2}{2T_S})&0\\
0&\tanh(\frac{\e\pm V/2}{2T_S})
\end{array}
\right)
\end{equation}
 The explicit form of the energy-dependent
functions $g_S^{R(A)},\;f_S^{R(A)}$ is:
\be
\label{defs}
 g_S^{R(A)}(\e)=\frac{\e}{\Delta}\left(\pm\eta_S-i\xi_S\right),\,\,
f_S^{R(A)}(\e)=\xi_S\pm i \eta_S,\;
\ee
where
\be
 \eta_S=\frac{\Delta\sign\e}{\sqrt{\e^2-\Delta^2}}\;\theta(|\e|-\Delta),\;
\xi_S= \frac{\Delta}{\sqrt{\Delta^2-\e^2}}\;\theta(\Delta-|\e|) \ee

\noindent {\bf 2. Spectral functions.} Here we  calculate Green
functions  $\hat G^{(R,A)}$ which describe proximity effect in
normal region. We write matrix Green function $\hat G^{R(A)}(x,\e,t)$ in the form \be
\label{green}
\hat G^{R(A)}=\left(\begin{array}{cc}\pm (1-g_1^{R(A)})&f_1^{R(A)}\\
f_2^{R(A)}&\mp(1-g_2^{R(A)})\end{array}\right)
\ee
Since the
proximity effect is weak due to the presence of barriers, we can expand
the equations for $G^{(R,A)}$ in powers of small parameter $r^{-1}$.
To the first order one finds:
\be
\begin{array}{c}
g_{1,2}^R\sim r^{-2},\;\; f_{1,2}^{R,A}\sim r^{-1}
\end{array}
\ee Besides, we assume here that $\varphi_-\sim r^{-2}$, which is
self-consistent assumption. Smallness of $\varphi$ is due to the
fact that almost all voltage drops at the barriers at $r \gg 1$. We
consider Eqs.(\ref{UsadelRA},\ref{KupriyanovRA})
  and keep terms, proportional to $r^{-1}$ in the equations, and terms of the order of unity in the
   boundary conditions for anomalous Green function $f_1^R(x,\e,t)$, to obtain:
\be \label{RALinearized}
\begin{array}{c}
 \p^2_x f_1^R+\kp^2
f_1^R=0 \\\\
\p_x f_1^R|_{x=\pm\frac12}=\pm r^{-1}e^{\pm
iVt}f_{S}^R
\end{array}
\ee Here \be\label{kappa}\kp_\e=\sqrt{\frac{2i\e}{\Th}-\gamma}\ee
 and $\g=(\tau_{in}\Th)^{-1}$ is the dimensionless inelastic scattering
rate. Linear approximation leading to Eqs.(\ref{RALinearized})
is applicable at all energies $\e$, if $(\g r)^{-1} \sim \tau_{in}E_g \ll 1$
(here $E_g$  is the proximity-induced minigap~\cite{McMillan} at $\g=0$).
Other Green functions are expressed via relations
\be
\begin{array}{c}f_{1,2}^A(x,\e,t)=f_{1,2}^R(x,-\e,t),\\\\
f_2^{R(A)}(x,\e,t)=f_1^{R(A)}(x,\e,-t)
\end{array}
\ee
The above equations are easy to solve, with the following
result:
\begin{small}
\be f_1^{R}=v^R \left[e^{iVt}\cos\kp_\e\left(x+\frac{1}{2}\right)
+e^{-iVt}\cos\kp_\e\left(x-\frac{1}{2}\right) \right] 
\label{spectral}
\ee
\end{small}
For the future convenience, we introduce definitions
\be
\begin{array}{c}
u^R=-\frac{f_S^R(\e)}{r}u(\e),\;\;v^R=-\frac{f_S^R(\e)}{r}v(\e)\\
u^A=-\frac{f_S^A(\e)}{r}u(-\e),\;\;v^A=-\frac{f_S^A(\e)}{r}v(-\e)
\end{array}
\ee with
$u(\e)=\frac{\cos\kp_\e}{\kp_\e\sin\kp_\e},\;\;v(\e)=\frac{1}{\kp_\e\sin\kp_\e}$\,
, so that 
\be f_{1,2}^{R(A)}(x=\frac{1}{2})=u^{R(A)}e^{\pm
iVt}+v^{R(A)}e^{\mp iVt}. 
\label{spectral2}
\ee 
Retarded and advanced normal Green
functions can be found from the normalization condition:
 \be
 g_1^{R}=\frac{1}{2}f_1^R\circ f_2^R,\quad g_2^R=\frac{1}{2}f_2^R \circ f_1^R
 \ee
 Electric potential is
\be
\label{electric}
\varphi(t)=\frac{1}{2}\int\left(h_3+g_\varphi^R\circ h_0-h_0\circ
g_\varphi^A\right) d\e
\ee
 with $g_\varphi^R=\frac{1}{8}\left[f_2^R\;\circ,\;f_1^R\right]$.
Linearization over $f_{1,2}$  breaks down at energies too close to the
 gap $\Delta$: $|\e-\Delta|\le\frac{\Th}{r^2}\max({\Th}/{\Delta},1)$, but the resulting
square-root singularity  $|\e-\Delta|^{-1/2}$ does not influence our
results.


{\bf 3. Kinetic equation and its solution.} In this section we
determine the electron distribution function. Its form is determined
by interplay of several processes. The electrons diffuse from one
superconducting lead to another during characteristic time
$\tau_D=\Th^{-1}$, and are subjected to normal and Andreev
scattering at the boundaries. Electron-electron scattering tends to
thermalize them to equilibrium with some effective temperature
$\Teff$. Electron-phonon scattering tends to bring $\Teff$ closer to
the substrate temperature $T_S$, thus taking the energy out of
electron system. Another (and more effective at low temperatures)
channel of electron cooling is the tunneling of hot electrons to
superconducting reservoirs~\cite{Courtois}. In the SINIS junction,
the role of inelastic scattering is relatively large, since Andreev
reflections are suppressed due to weakness of the proximity effect,
so that electron spends a long time, bouncing back and forth between
the boundaries. At $\chi= \g r^2\gg 1$ the electron distribution
thermalizes and reads \be\label{therm}\hat
h(x,\e,t)=h_0^{(0)}(\e)\hat\tau^0+O(\chi^{-1})\ee where
$h_0^{(0)}(\e)=\tanh(\frac{\e}{2\Teff})$ is equilibrium distribution
with some effective temperature $\Teff$; in general,  $\Teff\ne
T_S$. Effective temperature $\Teff$ has to be determined from the heat balance equation (cf. e.g. \cite{Raja}).
After that is done, one can calculate non-equilibrium correction to
the distribution function (the second term in (\ref{therm})). Note,
that it can be important in terms of calculating the Josephson
current, even if it small. The point is that thermal distribution
leads (see below) to the amplitude $|I_s|$ of the {\emph ac}
Josephson current $I(t)$, which is exponentially small in $L/\xi_T$.
 On the other hand, non-equilibrium corrections decay with temperature and length
 much slower. These non-equilibrium corrections are rather interesting,
since they arise due to coherent Andreev reflections.

In order to find this nonequilibrium correction, we simplify general
kinetic equations (\ref{UsadelK}), (\ref{KupriyanovK}) by separating
terms of different orders over small parameter $r^{-1}$, and
adopting the simplified form of collision integral, i.e.
"$\tau$-approximation". In doing so, one has to consider the
boundary condition (\ref{KupriyanovK}) up to the $r^{-1}$ terms  and
kinetic equation (\ref{UsadelK}) up to the $r^{-2}$ terms. Besides,
these are small energies $\e\ll\Delta$, where deviations from the
equilibrium are important, so we put $\eta_S(\e)=0,\;\xi_S(\e)=1$.
We look for the distribution function in the following form:

\be
\begin{array}{c}
h_0(x,\e,t)=h_0^{(0)}(\e)+r^{-2} h_0^{(2)}(\e,x,t)\\\\
h_3(x,\e,t)=r^{-2} h_3^{(2)}(\e,x,t)
\end{array}
 \ee

 Next, due
to the spatial symmetry, $h_3^{(2)}$ and $h_0^{(2)}$ are odd and even functions
of $x$, correspondingly. Taking this into account, we write boundary
conditions (\ref{KupriyanovK}) only at the $x=\frac{1}{2}$ boundary,
to obtain with the necessary accuracy: \be \label{KupriyanovKSimpl}
4\p_x h_{0,3}^{(2)}|_{x=\frac{1}{2}}=J_1\mp J_2
\ee
with $J_{1,2}$ terms which depend on $h_0^{(0)}$ only, and are
known therefore:
 \be
  J_{1,2}=r\left(f_{1,2}^R\circ e^{\mp iVt}X + e^{\pm iVt}X\circ f_{2,1}^A\right)\\\\
 \ee
 with $X(\e)=\frac12\left(h_{0+}^{(0)}-h_{0-}^{(0)}\right)$.

We used here special notation for the "energy-shift" subscript: for
any function of energy $f(\e)$ we define
\be
f_\pm(\e)=f(\e\pm\frac{V}{2}) \, \quad f_{++}=f(\e+V)\, \quad
f_{--}=f(\e-V) \, .
\label{shift}
\ee
Since the boundary conditions are known, we proceed with the kinetic
equation. Weakness of the proximity effect allows to neglect
modifications of the diffusion coefficient, and also the terms, mixing $h_0$ and
$h_3$ in the matrix equation (\ref{UsadelK}). Adopting $\tau$-approximation
for the collision integral, we obtain
\be
\label{UsadelKSimpl}
\begin{array}{c}
\Th\p_x^2 h_0^{(2)}-(\p_t h_0^{(2)}+\tau_{in}^{-1} h_0^{(2)})=0
\\\\
\Th\p_x^2 h_3^{(2)}-(\p_t h_3^{(2)}+\tau_{in}^{-1} h_3^{(2)} +
ir^2\varphi_- h_0^{(0)})=0
\end{array}
\ee
Kinetic equations ($\ref{UsadelKSimpl}$) are to be written
for each harmonic (in terms of the total time $t$) of the
distribution function, with the electric potential term that couples
$h_0^{(2)}$ and $h_3^{(2)}$ due to the self-consistency condition
(see (\ref{electric})). In the limit of large temperatures $\Teff\gg\Th$,
when the nonequilibrium contribution to the current becomes important,
this term is exponentially small in $L/\xi_T$ and we neglect it.
Besides, if $V\ll\Teff$, one has $X={V}/{4\Teff}$.

We look for the
nonequilibrium correction in the following form: \be
\begin{array}{c}h_0^{(2)}(x,\e,t)=
\sum_{n=0,\pm 1}
A_{n,\e}\cos (\kp_{nV}x)e^{-2iVn t}\\\\
h_3^{(2)}(x,\e,t)=\sum_{n=0,\pm 1} B_{n,\e}\sin
(\kp_{nV}x)e^{-2iVn t}
\end{array}
\ee

 Boundary conditions
(\ref{KupriyanovKSimpl}) allow to calculate $A,\;B$:
\be
\begin{array}{c}
A_{n,\e}=-\frac{r}{16}\frac{V}{\Teff}\frac{\tilde A_{n,\e}}{\kp_{nV}\sin(\frac12\kp_{nV})}\\\\
B_{n,\e}=\frac{r}{16}\frac{V}{\Teff}\frac{\tilde B_{n,\e}}{\kp_{nV}\cos(\frac12\kp_{nV})}
\end{array}
\ee
 with
\begin{small}
\bea
\nonumber
\begin{array}{c}
\tilde A(n,\e)=\left(\alpha_+ -\alpha_-\right)\delta_{n,0}+ \left(v_+^R -v_-^A\right)\delta_{n,1}-\left(v_-^R-v_+^A\right)\delta_{n,-1}\\\\
\tilde B(n,\e)=\left(\alpha_++\alpha_-\right)\delta_{n,0}+ \left(v_+^R+v_-^A\right)\delta_{n,1}+\left(v_-^R+v_+^A\right)\delta_{n,-1}
\end{array}
\eea
\end{small}
where $\alpha=u^R+u^A$.

{\bf 4. Time-dependent current}\\

With the distribution functions (\ref{therm}) being determined, we turn
to the calculation of time-dependent electrical current. It is
convenient to calculate it in the vicinity of the right boundary,
with $\hat j^{K}$ in (\ref{Current}).
There are lot of terms in
(\ref{Current}), but we keep only those of them, that are
proportional to the first order contribution to the anomalous
function $\hat G^{R(A)}$ and contribute to the oscillating part of the
current. This allows us to get the leading terms at
both low and high effective temperatures comparative to $\Th$. More
precise treatment will give some corrections, that are of the higher
orders of $r^{-1}$ at low temperatures and exponentially suppressed
at high temperatures. In the limit of large superconducting gap
$\Delta\gg\max(\Th,V)$ the result reads
\begin{small}
\be
\label{I1}
I(t)=\frac{1}{8R}\int\left[I_\e^{+}(t)-I^{-}_\e(t)\right]d\e
\ee
\end{small}
with \be I_\e^{\pm}(t)=(K_{1,2}\circ e^{\mp iVt}+e^{\mp iVt}\circ
K_{1,2}) \ee with $R$ being the resistance of the SINIS junction in
the normal state: $R=2R_B$. We have introduced here
 $K_{1,2}=f_{1,2}^R\circ h_2-h_1\circ f_{1,2}^A$, where anomalous and distribution functions are
 supposed to be calculated at $x=\frac{1}{2}$. To proceed, we note that $I(t)$ it can be considered as a sum of two
different contributions, according to (\ref{therm}). \\
{\bf\em Equilibrium current}. It reads: $I_{eq}=\Re[e^{-2iVt}I_1]$ with
 \be
\label{Is0}
I_1=-\frac{1}{2Rr}\int\left[v(\e+V/2)-v(-\e+V/2)\right]\tanh\frac{\e}{2\Teff}
d\e
\ee
We note that at zero-voltage limit one has
$h_{0}^{(0)}=\tanh(\frac{\e}{2T_S})$ and for the amplitude $I_{1}$
we get usual result~\cite{LarkinAslamazov} for the equilibrium critical current:
$I_1(V=0)= -\frac{i}{Rr}\int h_S\Im v\;d\e $.
In general, the integral (\ref{Is0}) can be reduced to the sum over residues to give
$I_1=-\frac{2\pi i\Teff}{Rr}\sum_{n=1}^\infty \frac{1}{q_n\sinh q_n}$, where
$q_n=\sqrt{\frac{2(2n-1)\pi \Teff-iV}{\Th}+\g}$.

In the limit of $\Teff\gg\Th$ and $\g \le 1$ we obtain
\be |I_1| =
c\frac{\Teff}{rR\sqrt{a}}e^{-\sqrt{a}} \left\{
\begin{array}{l}
a = \frac{2\pi\Teff}{\Th}\,\; c=4\pi, \quad V\ll\Teff \\\\
a = \frac{V}{2\Th}\,\; c=2\sqrt2\pi, \quad V \gg \Teff
\end{array}
\right.
\label{Is}
\ee
At very low temperatures
$\tanh\frac{\e}{2\Teff} \to \sign\e$ and the integral for the
current can be done explicitly:
\be |I_1| = \frac{2\Th}{rR}\left
|\ln\cot\left(\frac12\sqrt{\frac{iV}{\Th} - \gamma}\right) \right|
\label{Is1}
\ee
The above expression is valid for any voltage if $T\tau_{in} \ll 1$
and for high voltages $V\tau_{in} \gg 1$ otherwise.

{\bf\em Nonequilibrium current.} Calculation starting from
Eq.(\ref{I1}) leads to: \be
I_{neq}=\Re[e^{-2iVt}\Phi_1+e^{-4iVt}\Phi_2] \label{neq} \ee with
complex amplitudes $\Phi_{1,2} =
-\frac{\pi\Th}{16Rr^3}\frac{V}{\Teff}\phi_{1,2}$, where $\phi_{1,2}$
are given by:
\begin{small}
\be
\begin{array}{c}
\phi_1=\G_0(x_0-y_0)-\G_V(x_0+y_0)+(x_V+y_V)(\G_{2V}-\G_{V})
\\ \\
\phi_2=-(x_V+y_V)\G_{2V}
\end{array}
\label{neq0}
\ee
\end{small}
with $x_{\e}=-\frac{\cot(\kp_{\e}/2)}{\kp_{\e}},\;y_\e=\frac{\tan(\kp_{\e}/2)}{\kp_{\e}}$
and $\G_\e=\frac{1}{\sqrt{2(i\e/\Th-\g)}\sin\sqrt{2(i\e/\Th-\g)}}$.
The results (\ref{neq0}) are applicable for any values of $V/\Th$ ratio,
but in the lowest order over $V/\Teff$.

Here we analyze the case of weak inelastic scattering $\g \ll 1$. In the limit of zero voltage,
$|\phi_2|=2\g^{-1}|\phi_1|\gg|\phi_1|$ and the second harmonic dominates.
For small voltages $V\ll\Th$ one has:
\be
\Phi_1= -\frac{\pi\Th}{16Rr^3}\frac{V}{\Teff}\left\{
\begin{array}{l}
\frac12 \g^{-1}+4\g^{-4}(V/\Th)^2,\quad V \ll \tau_{in}^{-1}\\\\
-\g^{-2},\;\; \quad \tau_{in}^{-1}\ll V
\end{array}
\right.
\label{neq1}
\ee
and
\be
\Phi_2= -\frac{\pi\Th}{16Rr^3}\frac{V}{\Teff}\left\{
\begin{array}{l}
\g^{-2},\;\; \quad V\ll\tau_{in}^{-1}\\\\
-\frac{1}{4}(\Th/V)^2,\;\; \quad \tau_{in}^{-1}\ll V
\end{array}
\right. 
\label{neq2} 
\ee 
Comparing first lines of Eqs.(\ref{neq2})
and (\ref{Is}) we find that non-equilibrium second harmonics
dominates the {\em ac} current at $\Teff \geq \Th\ln^2(\gamma r)$
and low voltages $V \ll \tau^{-1}_{in}$. Similar phenomenon was
observed in Refs.~\cite{Courtois2001,Kroemer}, where subharmonic
Shapiro steps with slowly decreasing (upon $T$ increasing) amplitudes
were found.  Qualitative theory of this phenomenon was proposed by
Argaman~\cite{Argaman}, who discussed it in terms of time-dependent
Andreev bound states with non-equilibrium population. Our result
(\ref{neq2})  contains the same $V/\Teff$ dependence  at low $V$ as 
found in Ref.~\cite{Argaman}; 
however, we got $I_{neq} \sim \chi^{-1}I_{neq}^{\rm Argaman}$.
We expect therefore that the result of ~\cite{Argaman} is valid (up to numerical
factor of order unity) under the condition $\chi \leq 1$, which we do not 
consider here.
To understand the origin of the whole effect  it is useful to note 
that the result~\cite{Argaman} for the second harmonics  
is very similar to  (9) of Ref.~\cite{Zhou} where Debye relaxation contribution to 
the {\it dc}  conductance of SNS junction was estimated. Moreover, preprint version of
Ref.~\cite{Zhou} contains the same kind of estimation for the SINIS junction, with the
result $\propto \tau^2_{in}$, very much like our expression for $\Phi_2$.
We believe therefore that non-equilibrium second harmonics of the current
originates from the same Debye relaxation mechanism.

Non-equilibrium {\emph ac} current beyond linear in $V$
approximation never was  calculated previously, to the best of our
knowledge. Eqs.(\ref{neq1},\ref{neq2}) demonstrate that at not very
low voltages non-equilibrium  first harmonics of the current $\Phi_1$ 
becomes comparable to  $\Phi_2$ and then exceeds it. 
Note also $\pi$-shift in phases of $\Phi_{1,2}$  with
growth of voltage.
Technically, $\Phi_1$ originates from the peaks in nonequilibrium stationary parts 
of distribution functions,  $h_{0,3}^{(2)}(\e)$  at $\e = \pm V/2$.
These peaks result from the modulation of the spectral density with the
Josephson frequency, cf. Eq.(\ref{spectral2}).
For the particular value of $\g=0.2$, the absolute values of amplitudes 
$\phi_{1,2}(V)$  at low voltages $V\leq \Th \ll \Teff$ are plotted in the Figure.
\begin{figure}
\includegraphics[width=0.95\columnwidth]{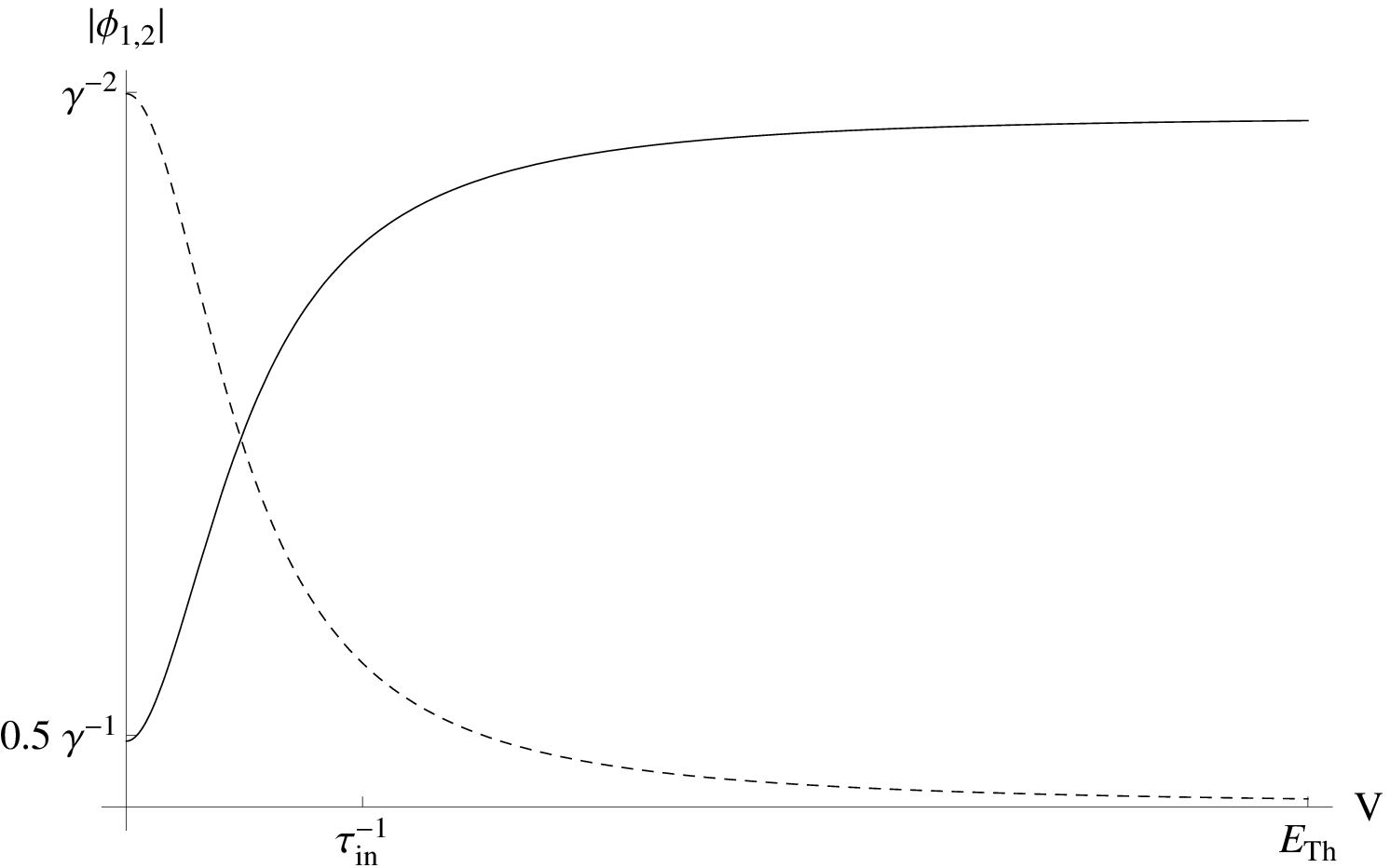}
\caption{\label{F:g_0} Nonequilibrium amplitudes $|\phi_{1}(V)|$ (normal line)
and $|\phi_{2}(V)|$ (dashed line) are
 calculated with Eq.(\ref{neq0}) for $\g=0.2$ and $V \ll \Teff$.}
\end{figure}

{\bf 5. Conclusions}\\

In this paper we have considered the AC Josephson effect in a long
SINIS junction in the case of
temperature and voltage low with respect to the bulk gap $\Delta$. The main results
 are given in Eqs.(\ref{neq0},\ref{neq1},\ref{neq2}) for non-equilibrium
contribution to the current in the high-temperature range $\Teff \gg
\Th$. At high electron temperatures $\Teff$ in the normal wire, the
nonequilibrium contribution to the current becomes dominant. We
found, that the {\emph ac} current contains two harmonics: the first with basic
Josephson frequency $\w_J = 2eV/\hbar$ and the second harmonic $2\w_J$.
Second harmonic has the largest amplitude at low voltage and high
temperature. Note, that calculation of non-equilibrium effects at
low temperatures is complicated due to the necessity to account for
the time-dependent electric potential in the normal wire; we leave
this problem for the future studies.

We are grateful to Ya. V. Fominov for many useful advises, and to H. Bouchiat,
H. Courtois, S. Gueron and V. V. Ryazanov for illuminating discussions.
This research was supported by RFBR grant 07-02-00310 and by the RAS Program
 "Quantum physics of condensed matter".

{\bf Appendix.}

Here we explicitly write out nonstationary Usadel equation and
boundary conditions. Retarded component of (\ref{Usadel}) takes the
following form: \be \label{UsadelRA} -D\p_x(\hat G^R\circ\p_x\hat
G^R)+\p_\tau[\hat \tau^3,\hat G^R]+\frac12\p_T\{\hat \tau^3,\hat
G^R\}+i\varphi_-\hat G^R=\hat I^R \ee with $\hat
I^R=-i\left[\hat\Sigma^R\;\circ,\;\hat G^R\right]$. It is
supplemented with retarded component of boundary condition
(\ref{Kupriyanov}): \be \label{KupriyanovRA} 2R_{SN}\sigma_N
j_l^R=[\hat G_l^R\circ, \hat G_r^R]. \ee Since advanced components
are identical (with $R\to A$ substitution), we proceed with kinetic
equation, which results from the Keldysh component of Usadel
equation:
\begin{small}
\be
\begin{array}{c}
\label{UsadelK} -D\left[\p_x\left(\p_x \hat h - \hat
G^R\circ\p_x\hat h\circ\hat G^A\right)+\left(\hat j^R\circ\p_x\hat
h-\p_x\hat h\circ\hat j^A\right)\right] \\\\
+\left(\hat G^R_+\circ\p_T\hat h+i\hat G^R\circ\varphi_-\hat
h\right)-\left(\p_T\hat h\circ\hat G^A_++i\varphi_-\hat h\circ\hat
G^A\right)=\hat I^{St}
 \end{array}
 \ee
\end{small} 
with
\be
\begin{array}{c}
 \hat
G^{R(A)}_+=\frac12\left\{\hat G^{R(A)},\hat\tau^3\right\},\;\;\hat
I^{St}=-i\left(\hat G^R\circ\hat\sigma-\hat\sigma\circ\hat
G^A\right)\\\\
\hat\sigma=\hat\Sigma^R\circ\hat h -\hat
h\circ\hat\Sigma^A-\hat\Sigma^K.
\end{array}
\ee
The Keldysh component of boundary conditions (\ref{Kupriyanov})
reads \be \label{KupriyanovK} 2R_{SN}\sigma_N\left(\p_x\hat h_l-\hat
G_l^R\circ\p_x\hat h_l\circ\hat G^A_l\right)=\left(\hat
G^R_l\circ\hat u-\hat u\circ\hat G^A_l \right) \ee
 where
$\hat u=\hat G^R_r\circ\delta\hat h-\delta\hat h\circ\hat
G^A_r,\;\;\;\delta\hat h=\hat h_r-\hat h_l$

\end{document}